\documentclass[page-classic]{epl2} 

\usepackage{amssymb,bm,amsfonts,amsmath,graphicx,color}
\title{Observation of low energy dispersive modes in underdoped 
\chem{(La,Nd)_{2-x}Sr_xCuO_4}}

\author{Matteo d'Astuto \inst{1} \and Sylvain Denis \inst{2} \and Jeff Graf \inst{3} \and Alessandra Lanzara \inst{3,4} \and Claudia Decorse \inst{2} \and Paola Giura \inst{1} \and Sang-Wook Cheong \inst{5} \and Abhay Shukla \inst{1} \and Patrick Berthet \inst{2} \and Daniel Lamago \inst{6,7} \and John-Paul Castellan \inst{6,7}  \and Alexei Bossak \inst{8} \and Michael Krisch \inst{8}}

\shortauthor{Matteo d'Astuto \etal}

\institute{                    
  \inst{1} Institut de Min\'eralogie et de Physique des Milieux Condens\'es (IMPMC), Universit\'e Pierre et Marie Curie - Paris 6, case 115, 4, place Jussieu, 75252 Paris cedex 05, France\\
  \inst{2} ICMMO, UMR 8182, Universit\'e Paris Sud and CNRS, 91405 Orsay, France \\
  \inst{3} Materials Sciences Division, Lawrence  Berkeley National Laboratory, Berkeley, CA 94720, USA\\
  \inst{4} Department of Physics, University of California Berkeley, CA 94720, USA\\
  \inst{5} Rutgers Center for Emergent Materials and Department of Physics \& Astronomy, Rutgers, Piscataway, New Jersey 08854, USA\\
  \inst{6} Institut f\"ur Festk\"orperphysik, Karlsruher Institut f\"ur Technologie, D-76021 Karlsruhe, Germany\\
  \inst{7} Laboratoire Leon Brillouin, CEA-Saclay, F-91191 Gif sur Yvette Cedex, France\\
  \inst{8} ESRF - The European Synchrotron, CS40220, 38043 Grenoble Cedex 9, France}


\date{\today}

\abstract{
We find excitations lower in energy than known phonon modes in underdoped La$_{2-x}$Sr$_x$CuO$_{4+\delta}$ (x=0.08), with both inelastic X-Ray scattering (IXS) and inelastic neutron scattering (INS). A non dispersive excitation at 9 meV is identified and is also seen by INS in (La,Nd)$_{2-x}$Sr$_x$CuO$_{4+\delta}$, with 40$\%$ Nd substitution. INS also identifies a still lower energy dispersive mode at low q in the Nd free sample. These modes are clearly distinct from the longitudinal acoustic phonon and correspond in energy to the Zone Centre modes measured by optical spectroscopy and associated with stripe dynamics.
}

\begin{document}

\maketitle

The relevance of exotic charge-modulation states and their co-existence with the superconducting state in underdoped high temperature superconductors is a subject of current debate.
This state has now been conclusively detected in a variety of materials where this ordering is \textit{static} and is linked to lattice deformations and strain \cite{bianconi1, tranquada1,orenstein,rmp-kievelson,bonn,abbamonte,ghiringhelli_ybco-stripes}.
Reports of static stripes exist also in materials like nickelates \cite{cheongPRL93, cheongPRB94,yoshizawa} where superconductivity is never attained while in
cuprates pinning of stripes by lattice deformations is accompanied by suppression of superconductivity \cite{tranquada1,orenstein,rmp-kievelson,bonn,abbamonte}. 
Available evidence of dynamical charge-modulations coexisting with superconductivity is mostly indirect  \cite{ando1,sun,niedermayer,bianconi2}.

A related debate concerns the interpretation of charge-modulations in cuprates as either stripes, \textit{i.e.}, driven by Coulomb repulsion \cite{tranquada1}, or as Charge-Density-Waves (CDW) either driven by lattice deformations as originally claimed \cite{bianconi2},  or, more recently, by a nesting of the Fermi Surface \cite{Wise:2008uq, ghiringhelli_ybco-stripes}. 
The two interpretations give completely different doping dependences of the charge modulation wave-vector: for the first one we expect an \textit{increase} with hole-doping \cite{tranquada1,orenstein,rmp-kievelson}, while for the CDW nesting-driven case we expect a \textit{decrease} with hole-doping  \cite{Wise:2008uq}. Both scenarios yield comparable wave-vectors close to x $\approx$ 1/8 where they are most easily detected \cite{dastuto-laba}.
Hence it is fundamental to characterize possible signatures of such charge-modulations across the entire phase diagram of cuprates.

Optical spectroscopies such as Raman and infrared experiments have revealed the presence of two new low energy modes in underdoped La$_{2-x}$Sr$_x$CuO$_{4+\delta}$ \cite{tassini, sugai, caprara, lucarelli, venturini}.  
The first mode occurs at 25 cm$^{-1}$ ($\sim$ 3 meV) at low temperature \cite{tassini, sugai, lucarelli} and shows a strong temperature dependence:  as the temperature increases the mode frequency increases up to $\approx$ 200 cm$^{-1}$ while its intensity decreases, eventually becoming zero above $\approx$ 200 K \cite{tassini}. 
The second mode occurs at higher energy, at about 80 cm$^{-1}$ ($\sim$ 10 meV) \cite{sugai} at low temperature. 
More recently, time-resolved optical spectroscopy in the femto-second regime showed a resonant mode of about 2.0 THz frequency that corresponds to the same energy range ($\approx$ 70 cm$^{-1}$ or  8 meV) \cite{Torchinsky:2013kx}. Such mode is interpreted as an amplitudon, in line with Ref. \cite{sugai}, suggesting a charge-density-wave like model. 

The interpretation of these Raman modes is the subject of debate:  do they result solely from scattering by electronic excitations \cite{tassini, caprara} or are they coupled charge-lattice modes \cite{sugai}? 

Recently IXS experiments have reported anomalies in low energy acoustic modes or additional modes at lower energies in cuprate Bi2201 \cite{bonnoit-bisco-ph} and oxychlorides \cite{dastuto-ccocoPRB}. 
In the "123" family (as YBCO, NdBCO,...)  acoustic phonon anomalies have been detected \cite{letacon-ybco-ph}, and linked to CDW \cite{letacon-ybco-ph,ghiringhelli_ybco-stripes}. However the charge modulations in the "123" family bear distinct feature from the one of the "214" family (\textit{e.g.}  \chem{(La,Nd)_{2-x}Sr_xCuO_4}) we study here. In particular the "123" family show no associated spin-fluctualtion \cite{ghiringhelli_ybco-stripes} in strike contrast to what observed in the  "214" one \cite{tranquada1,orenstein}. 

Here we report an observation of anomalous low energy modes for the first time in underdoped La$_{2-x}$Sr$_x$CuO$_{4+\delta}$ using both IXS as in the above cited studies, as well as INS, thus narrowing drastically the possible origins of the nature of the excitation and its dispersion.  We identify two additional modes at energies compatibles with the results obtained by optical spectroscopies\cite{tassini, sugai, lucarelli} at the zone centre, and extend the observation to the whole Brillouin Zone, as well as to the Nd substituted sample.
In the y=0.4 Nd substituted sample where the charge modulation is pinned \cite{tranquada1} we still observe a non dispersive mode while the lower energy dispersive mode is much weaker, and even not detected for reduced wave-vectors $(q, 0, 0)$ with 0.2 $\leq q \leq$ 0.4.
Using inelastic neutron scattering, we have explored the same (Q,E) region in a crystal with the same doping, x=0.08, in order to identify the possible origin of these excitations by comparing neutron and non-resonant X ray cross sections. We find strong evidence that they are coupled to lattice modes. 

 IXS and INS measurements were performed on two underdoped La$_{2-x}$Sr$_x$CuO$_{4+\delta}$ both with x=0.08, where Raman
measurements \cite{sugai} report the strongest intensities for the extra modes. The crystal for INS measurements (sample label S1) has cylindrical shape with an average diameter of  about 6 mm, a height in the order of of $\sim$ 35 mm and T$_c$=20 K, while on IXS experiments we measured a second crystal with a platelet shape, about 0.5 mm thick and few mm wide, with T$_c$=(20.7$\pm$0.2) K (sample label S2).
With INS we also measured a La$_{2-x-y}$Nd$_y$Sr$_x$CuO$_{4+\delta}$ crystal of cylindrical shape with an average diameter of about (5 $\pm$ 0.5) mm and a height of $\sim$ 25 mm grown from the melt in an image furnace by the traveling solvent floating zone method, with  with x=0.08 (as above) and y=0.4 (T$_c$=(7.5$\pm$0.25) K, sample label SNd).
 
The sample was mounted in reflection geometry on the cold finger of a closed-loop helium cryostat.  
The sample $c$ crystal axis was perpendicular to the scattering plane, with a beam spot of h $\times$ v $\approx$ 0.3 $\times$ 0.1 mm, while the cm$^3$ sample for the INS experiment was completely irrigated by the neutron beam. 
The highly collimated X-ray beam and small sample size result in a small probed volume and an angular width of $0.01^{\circ}$ FWHM for the (4 0 0) Bragg reflection. 
In the neutron experiment we measured  $1.5^{\circ}$ (S1) and $0.8^{\circ}$ (SNd) FWHM for the (2 0 0) reflection. Phonons were measured in the fourth Brillouin zone in the IXS experiment and from second to third zones by INS. 
The IXS spectrometer \cite{masciovecchio1,verbeni} was used with simultaneous measurements from 5 analyzers, and a resolution of $\approx 3 meV$   (see \textit{e.g.} Ref. \cite{mgb2-dastuto} for general details, and Ref. \cite{dastuto-laba} for the setup specific to the present experiment).  
We note that IXS spectra do not have contributions from incoherent scattering or multiple scattering to their $S(\mathbf{Q},\omega)$ dynamical
structure factor.  For X-rays incoherent Compton scattering occurs above $\approx$ 1 eV energy transfer due to the small electron mass, while multiple scattering is strongly diminished by the high photoelectric absorption cross section of cuprates. 
The multiphonon part is then typically the only  intrinsic background in IXS for meV energy transfers, and can be easily singled out \cite{baronprb}. 
On INS experiments we occasionally observe additional contributions in the measured energy range, mostly from higher monochromator harmonics. These can be calculated and thus easily eliminated.  
A standard lattice dynamical calculation using a shell model code \cite{mirone} was performed to calculate the phonon dispersion and the expected intensities from inelastic scattering as described in Ref. \cite{dastuto-laba}. 

\begin{figure*}[t]
\includegraphics[width=0.495\textwidth]{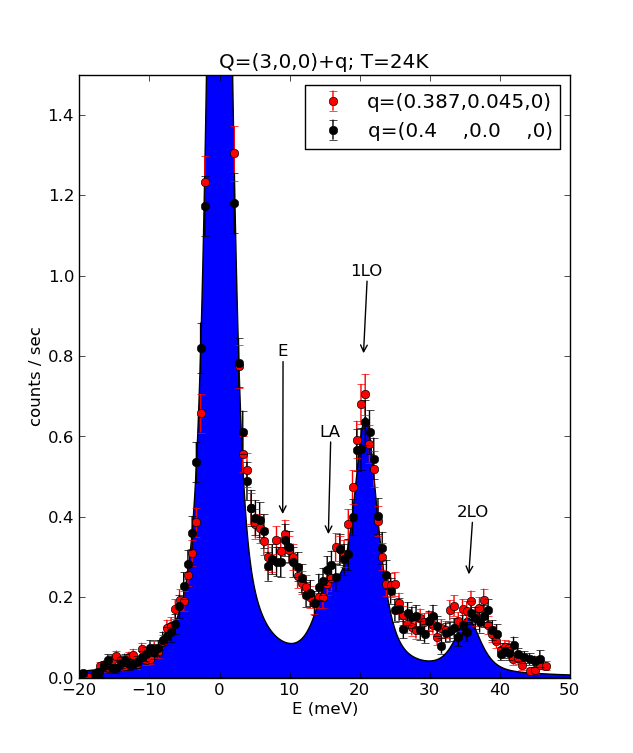} \includegraphics[width=0.495\textwidth]{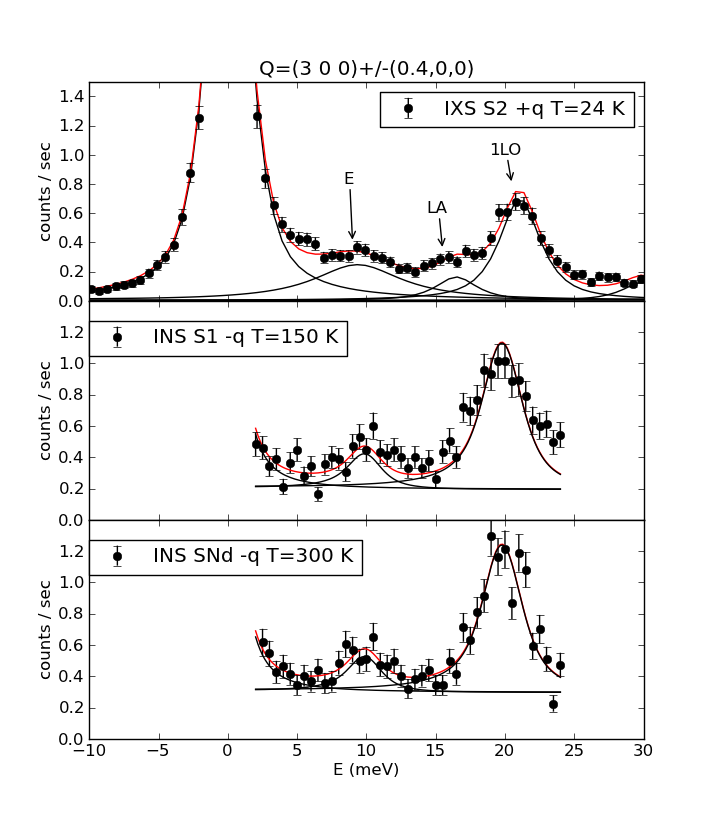}
\caption{\label{comp_calc_ixs_ins}(Color online). 
  Left panel. Energy loss IXS spectra in longitudinal
   configuration for $Q=(3+\xi, \upsilon, 0)$ at T = 24 K. Black circles:
   $\xi=0.387$ and $\upsilon=0.045$; red (gray) circles: $\xi=0.400$ and $\upsilon=0.0$. The blue (gray) filled area shows a calculation of the expected intensity from our model, integrated over the q-resolution volume around $Q=(3.4, 0, 0)$, convoluted to the instrumental energy resolution function. Right panel. Energy loss IXS and INS spectra and best fit to a sum of harmonic oscillators convoluted to the instrumental energy resolution function. The data are taken at  $Q=(3.4, 0, 0)$, for IXS, as in the left panel, and at the equivalent position $Q=(2.6, 0, 0)$ for INS data. The latter are taken for different temperatures and sample of  La$_{2-x-y}$Nd$_y$Sr$_x$CuO$_{4+\delta}$ with x=0.08 and different Nd substitution:  y=0 for sample S1 and S2 and y=0.4 for SNd.} 
 \end{figure*}

 
In the left panel of Fig. \ref{comp_calc_ixs_ins} we plot the IXS intensity as a function of energy for reduced wave-vectors $q=(0.387, 0.045, 0)$ 
and $q=(0.4, 0, 0)$. We detect three phonon modes which can be identified as the longitudinal acoustic mode (labeled LA, $\approx 16~ meV$), the first optical mode (labeled 1LO, $\approx$ 20 meV) and the second optical mode (2LO, $\approx$ 35 meV). 
These are expected from shell model calculations but also from earlier measurements \cite{pintrev1,fukuda}. 
We also observe an additional low energy mode (labelled E, $\approx 9~ meV$) which at this wave-vector point clearly emerges as a peak with intensity comparable to the other phonon modes.
The deviation from the longitudinal condition along $(1 0 0)$ for the $q=(0.387, 0.045, 0)$ spectra is negligible according to calculation, with less than 3\% transverse contribution. 
Indeed, the data shows no visible difference with the spectra measured in exact longitudinal condition with $q=(0.4, 0, 0)$.
We also check for contribution coming from other symmetries such as transverse modes. 
Though the measuring geometry used forbids 
the measurement of transverse phonons, the non-zero solid angle covered by the analyzer
may favor such a contribution. However in our case
the analyzer acceptance angle is too small to account
for such a parasitic signal from transverse phonons. 
In Fig. \ref{comp_calc_ixs_ins}, left panel, we show the calculated IXS intensity, integrated over the whole scattering volume in reciprocal space, according to our resolution and mosaic spread, and convoluted with the experimental energy resolution, by the blue filled area \footnote{A similar procedure has been used in Ref. \cite{PhysRevLett.113.025506}, and described in the supplemental material, Fig. FigSI2.}. The calculated intensity is  concentrated under the measured phonon modes, but there are no contribution from phonon modes to the scattered intensity $\approx 9 meV$. 
The right panel of Fig. \ref{comp_calc_ixs_ins} shows representative IXS and INS data at $q=(0.4, 0., 0)$ for temperatures ranging from 24 to 300 K, together with the corresponding fits using a model function as described elsewhere \cite{jeff-brief,dastuto-ccocoPRB}.  INS data are fitted to a damped harmonic oscillator convoluted with the instrumental function as calculated by a program including the spectrometer characteristics \cite{afitv}.
For all temperatures, we detect the phonon modes and the additional low energy excitation labelled "E" in Fig \ref{comp_calc_ixs_ins}, left panel.
A second mode, seen by Raman \cite{tassini} at lower energies, is also observed in the INS data, but only in a limited range of the Brillouin Zone, from $q=(0.07, 0, 0)$, where it intercepts the acoustic phonon mode dispersion, at approximately 5 meV, to $q=(0.4, 0, 0)$, where it seems to merge with the higher mode "E" at about 10 meV. 

\begin{figure}[h]
\includegraphics[width=0.45\textwidth]{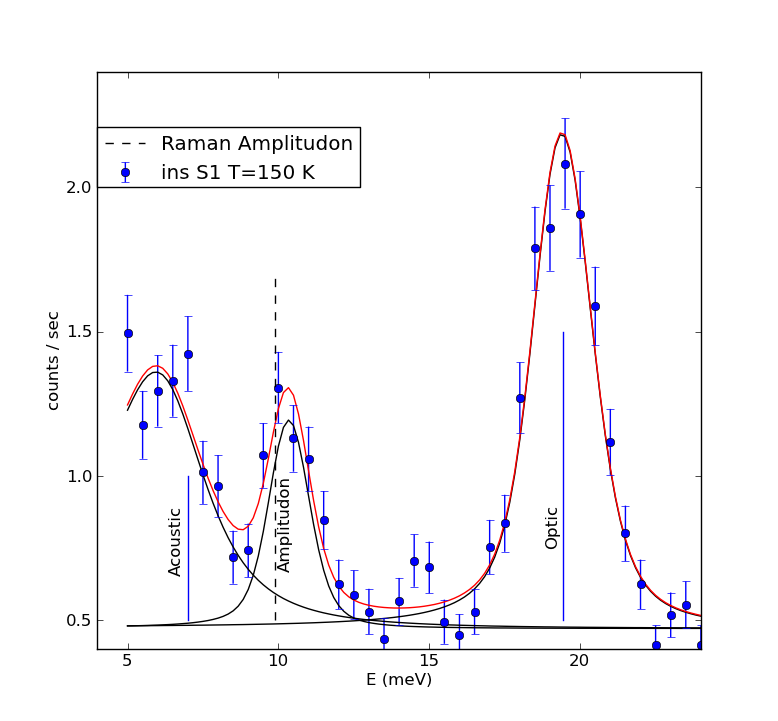}
\caption{\label{INS@BZC}(Color online). Raw INS energy scan on \chem{La_{2-x}Sr_xCuO_{4+\delta}} with x=0.08 (S1) at $Q=(3, 0, 0)$  and T=150 K. Red lines shows a best fit to a sum of harmonic oscillators convoluted to the instrumental energy resolution function, black lines representing each harmonic oscillators contribution (A single black line  when only one contribution is fitted).  Blue vertical line indicates the calculated position for the acoustic (low energy) and first optic (high energy) mode, while the black, dashed line indicate the position of the \textit{"Amplitudon"} according to Raman data in Ref. \cite{sugai}. 
 }
 \end{figure}

Both for IXS and INS we do not have low energy data close to the Bragg reflection. However, we have zone centre data around forbidden reflections $(2H+1 ,0 , 0 )$. We represent our data in an extended Brillouin Zone, where point with reduced wav-vector $(q,0,0)$ for $q \geq 0.5$ are unfolded up to these forbidden reflections, as usually done for I4/mmm-like symmetries \cite{reznik-rev}. But these points in fact can be folded back to the Zone Centre, and have been measured in the INS experiment, as shown in Fig. \ref{INS@BZC}, and can be directly compared to the optical results. In that figure we show indeed how the energy found in the Raman results of Ref. \cite{sugai} directly compare to our INS data, giving strong evidence that we are actually measuring the same feature. Unfortunately, such measurements are not possible with IXS, because of the presence of a strong parasitic elastic scattering, while the INS measurements are limited to E $\geq 5$ meV, due to geometrical constraint at large exchanged wave-vector, so that we can access only the excitation E, identified as an amplitudon mode in \textit{e.g.} Ref. \cite{sugai,Torchinsky:2013kx}, while the feature at low energy can not be measured at the zone centre. 


\begin{figure*}[t]
\includegraphics[width=0.495\textwidth]{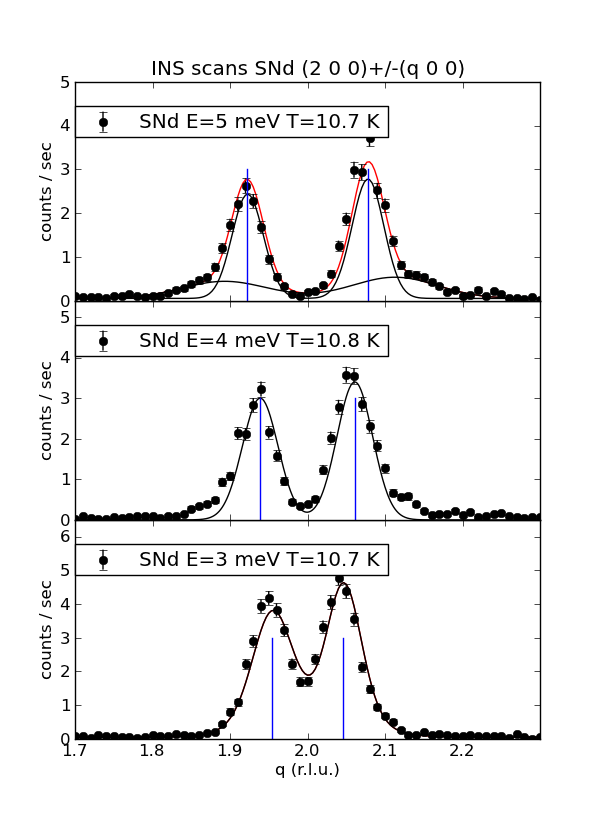}\includegraphics[width=0.495\textwidth]{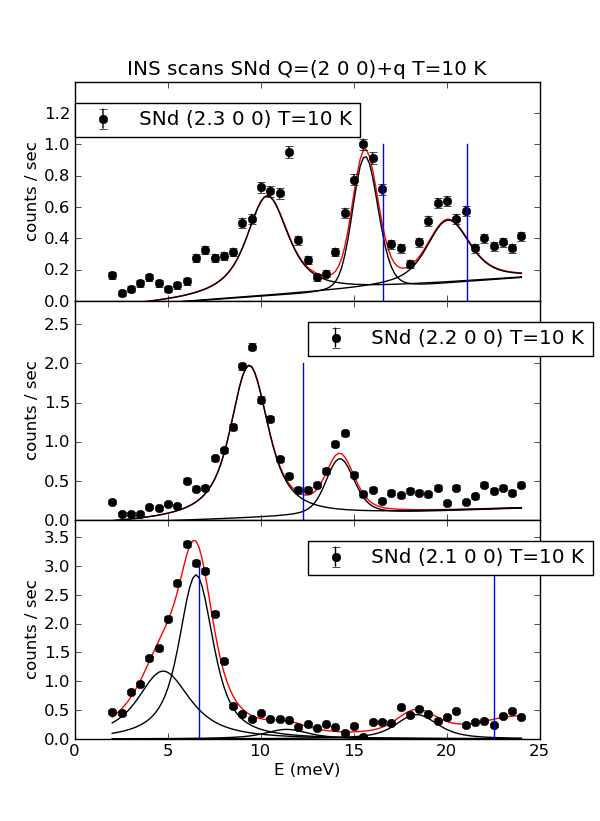}
\caption{\label{comp_ins_scan}(Color online). Raw INS data of La$_{2-x-y}$Nd$_y$Sr$_x$CuO$_{4+\delta}$ with x=0.08 and y=0.4 (SNd) along $Q=(2+q, 0, 0)$  at $\approx$ 10 K. Left panel: fixed-energy, $q$-scans, close to the Zone Centre $(2, 0, 0)$. Right panel:  fixed-$q$ energy scans. Red lines shows the best fit to a sum of harmonic oscillators convoluted with the instrumental energy resolution function; black lines represent individual harmonic oscillator contributions (A single black line  when only one contribution is fitted).  Blue vertical dashes indicate the calculated position for the acoustic mode at $\pm q$ in the left panel, and acoustic (low energy) and first optic (high energy) mode in the right panel. 
 }
 \end{figure*}


In Fig. \ref{comp_ins_scan} we compare INS intensities for several fixed-energy $q$-scans (close to the Brillouin Zone Centre) and fixed-$Q$ energy scans on the same sample at low temperature (10 K) with phonon modes expected from calculations. 
For energy losses below 5meV in the $q$-scans of the left panel only the acoustic phonon is seen. The energy scans of the right panel clearly show a strong anomalous mode at $\approx 9 meV$.

\begin{figure}
\includegraphics[width=0.8\textwidth]{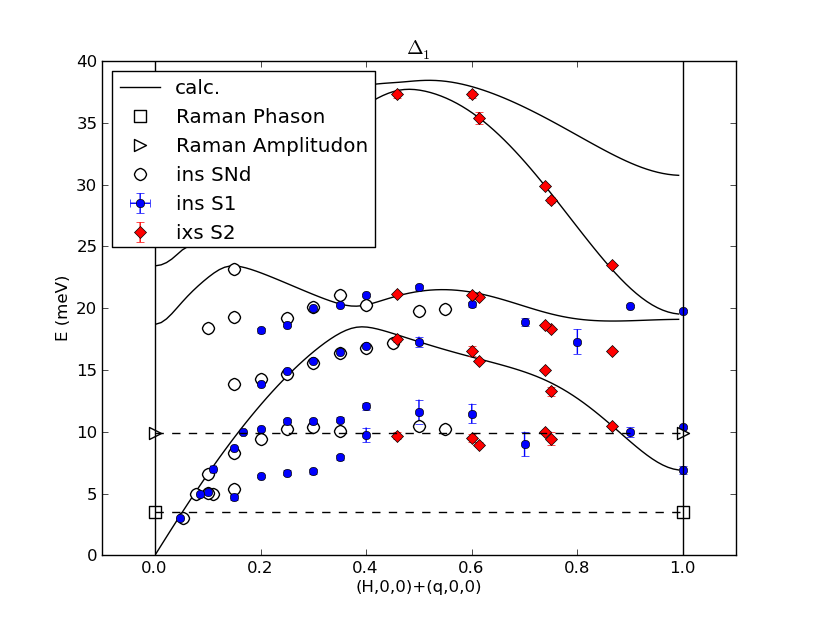}
 \caption{\label{disp}
 (Color Online).
 Dispersion of the longitudinal phonon modes with $\Delta$ symmetry along (3+$\xi$ 0,0). Note that this is an extended Brillouin Zone, with the zone centre at (2H,0,0) and zone boundary at (2H+1, 0, 0) corresponding to the  \textit{in-plane} (2$\pi$/a , 0) point.    
 Calculations (lines) are compared to the experimental results, as extracted from IXS (red filled diamonds) and INS (blue and white filled circles) spectra.  
 Data are shown up to the 2LO mode, the complete set of optic modes is reported in Ref. \cite{reznik-rev}. 
  INS data are reported for two different Nd substitutions as in Fig.  \ref{comp_calc_ixs_ins}. Raman data are taken from Ref. \cite{sugai}, and we trace a dashed line connecting these data at the Brillouin Zone Centre with the same ones unfolded at the extended zone boundary.}
 \end{figure}

In Fig. \ref{disp} we show the experimental and calculated dispersion of phonons in the extended zone in the usual representation for these systems \cite{reznik-rev}. 
The experimental dispersion is obtained from that of the fitted peaks in the experimental energy loss spectra as shown in Fig. \ref{comp_calc_ixs_ins} and \ref{comp_ins_scan}.
Phonon energies and their dispersion correspond well to the calculation in the second half of the extended Brillouin zone, from ($\pi$/a, 0 0)  to  (2$\pi$/a, 0 0) while the extra mode already shown in Fig. \ref{comp_calc_ixs_ins} and \ref{comp_ins_scan} now shows up as a flat, optical-like band. In the first half of the Brillouin zone this mode interacts strongly with the acoustic phonon opening a gap as in an avoided crossing (or anti-crossing) (see middle figure of the right panel of Fig. \ref{comp_ins_scan}). A lower energy dispersive component also appears which increases in energy with $q$, finally merging into the higher energy anomalous mode at $q=(0.4, 0, 0)$.

The INS measurements confirm the observations made using IXS and we clearly detect the additional mode at approximately 9-10 meV. Moreover we observe its interaction with the acoustical phonon as well as another anomalous dispersive mode at lower energy. The IXS and INS frequencies superpose well in the reduced q-range where we have access to both. This additional mode is consistently seen with different experimental set-ups and techniques, over different Brillouin zones, and on samples with different mosaic.
This has several important implications: 
\begin{itemize}
\item[a)]	Any possible instrumental artefact is excluded;
\item[b)]	The anomalous modes are linked to the periodicity of the crystal since measurements were made in different Brillouin zones;
\item[c)]	The mode "E" implies a displacement of the centre of the mass of the atoms, as it is seen both in the INS and IXS channel. In any case, a pure charge (in principle visible only by IXS) or spin (visible only by INS) excitation is excluded.
\end{itemize}

In fact, if this peak was of purely electronic origin we would not have been able to detect it with INS, and even for IXS the corresponding cross section would be several orders of magnitude smaller than that of the low energy phonon modes. 

We also confirm that the additional modes appear only along (1 0 0), but not along the diagonal (1 1 0) nor in transverse configuration, where the data match previously reported experiments. 
In addition, we observe a clear anti-crossing with the acoustic longitudinal mode of $\Delta 1$ character, indicating that the LA and the "E" mode possess the same symmetry.

From a phenomenological point of view, parallels can be drawn between the symmetry breaking associated with stripe formation and that coming from the formation of charge density waves. This point has been recently studied by optical, time-resolved experiment \cite{Torchinsky:2013kx} in the same system, as well as in a very similar experiment in YBCO \cite{letacon-ybco-ph}. 
In the latter case, however, a true static lattice deformation accompanies the electronic transition and has a drastic
effect on the phonon spectrum: softening of phonon branches to the point where the static lattice symmetry is altered. 
This softening and the new symmetry generate two new modes: the acoustic like phason with a very rapid dispersion given by the softening, and the optical like
amplitudon \cite{ravi}. In our case we do not observe similar softening, associated to a possible "phason" mode, even when we add Nd, that is supposed to pin the "stripe" order. On the one hand the effects may be subtle since fluctuating stripes would not bring about a change in the static lattice symmetry. On the other at a Sr-doping $x$ quite far from 1/8, the pinning effect could be weak. 

Our measurements point to a local symmetry breaking on the time-scale corresponding to THz frequencies, resulting from the coupling of the lattice with spatial charge modulations, possibly with dynamical and short range character.
They correspond favorably with the Brillouin Zone Centre measurements by Raman and infrared experiments for a similar doping, where they are identified as due to stripes.
An important open question has been the possible link between this phase and the lattice.  We show here that the low energy lattice dynamics bears an observable
fingerprint of this phase. 
Further studies close to the x $\approx$ 1/8 doping where the Nd substitution pins the charge-modulations as well as in the tetragonal phase far from structural transition will allow us to disentangle possible lattice effects from purely electronic ones. 


\acknowledgments
We thank Genda Gu for providing us the crystal (S1).
We are very grateful to R. Hackl, S. Lupi and E. Cappelluti for useful discussion and D. Gambetti for technical help. 
This work was supported by ESRF through experiment HS-2440 and HS-2689 and by
the U.S. Department of Energy under Contract No. DE-AC03-76SF00098, and LLB through experiment N. 11234. The IXS and INS work was partially supported by Berkeley Labs programs on "Quantum Materials" funded by the US Department of Energy, Office of Science, Office of Basic Energy Sciences, Material.
We acknowledge the support of the National Science Foundation through Grant No. DMR-0349361 and DMR-0405682, as well as the support from University of California, Berkeley, through France Berkeley Fund Grant.


\end{document}